\documentclass[a4paper,times,10pt,3p]{elsarticle}
  \usepackage{graphics}
  \graphicspath{ {./FIG/} }
 \usepackage{graphicx, amstext, amssymb, amsfonts, amsmath}
 \RequirePackage[colorlinks=true,citecolor=blue,urlcolor=blue,linkcolor=blue]{hyperref}
 \usepackage[usenames]{color}
 \definecolor{dark-green}{RGB}{0,100,0}
 \usepackage{bm}
 
 \newcommand{\comu}[1]{\left[#1\right]}

\makeatletter
\def\ps@pprintTitle{%
 \let\@oddhead\@empty
 \let\@evenhead\@empty
 \def\@oddfoot{}%
 \let\@evenfoot\@oddfoot}
\makeatother

\begin{document}
\begin{frontmatter}
\title{Generalization of the possible algebraic basis of $q$-triplets\footnote{To appear in European Physical Journal Special Topics.}}
\author[cbpfsf]{Constantino Tsallis}
\ead{tsallis@cbpf.br}
\address[cbpfsf]{
Centro Brasileiro de Pesquisas Fisicas and National Institute of Science and Technology for Complex Systems, Rua Xavier Sigaud 150, 22290-180 Rio de Janeiro-RJ, Brazil, and
Santa Fe Institute, 1399 Hyde Park Road, Santa Fe, NM 87501, USA }
\begin{abstract}
The so called $q$-triplets were conjectured in 2004 [Tsallis, Physica A {\bf 340}, 1 (2004)] and then found in nature in 2005 [Burlaga and Vinas, Physica A {\bf 356}, 375 (2005)]. A relevant further step was achieved in 2005 [Tsallis, Gell-Mann and Sato, PNAS {\bf 102}, 15377 (2005)] when the possibility was advanced that they could reflect an entire infinite algebra based on combinations of the self-dual relations $q \to 2-q$ ({\it additive duality}) and $q \to 1/q$ ({\it multiplicative duality}).  The entire algebra collapses into the single fixed point $q=1$, corresponding to the Boltzmann-Gibbs entropy and statistical mechanics. For $q \ne 1$, an infinite set of indices $q$ appears, corresponding in principle to an infinite number of physical properties of a given complex system describable in terms of the so called $q$-statistics. The basic idea that is put forward is that, for a given universality class of systems, a small number (typically one or two) of independent $q$ indices exist, the infinite others being obtained from these few ones by simply using the relations of the algebra. The $q$-triplets appear to constitute a few central elements of the algebra. During the last decade, an impressive amount of $q$-triplets have been exhibited in analytical, computational, experimental and observational results in natural, artificial and social systems. Some of them do satisfy the available algebra constructed solely with the additive and multiplicative dualities, but some others seem to violate it. In the present work we generalize those two dualities with the hope that a wider set of systems can be handled within. The basis of the generalization is given by the {\it selfdual} relation  $q \to q_a(q) \equiv \frac{(a+2) -aq}{a-(a-2)q} \,\, (a \in {\cal R})$. We verify that $q_a(1)=1$, and that $q_2(q)=2-q$ and $q_0(q)=1/q$. To physically motivate this generalization, we briefly review illustrative applications of $q$-statistics, in order to exhibit possible candidates where the present generalized algebras could be useful. 
\end{abstract}\end{frontmatter}

\section{Introduction} 

The goal of statistical mechanics is, starting from 
the microscopic natural rules (classical, relativistic, quantum mechanics, chromodynamics) and adequately using
probability theory, to arrive to the thermodynamical relations. Along these connections between the macro- and micro- worlds, the ultimate link is made through the fundamental concept of~\emph{entropy}.
This finding, accomplished against a stream of criticism, surely is one of
the most powerful and fruitful breakthroughs of the history of physical sciences. It was achieved by Boltzmann in the last three decades of the nineteenth century.
His result, currently known by every pure and applied scientist,  
and carved on his tombstone in Vienna, namely,
\begin{equation}
S_{BG} = k\ln W\,,
\label{eq:Tombstone}
\end{equation}
is the mathematical link between the microscopically fine description (represented by~$W$, the total number of accessible microscopic states of the system)
and the macroscopic measurable quantities (represented by the entropy $S_{BG}$, the very same quantity introduced by Clausius in order to complete thermodynamics!).
Apparently, Eq.~\eqref{eq:Tombstone} has been explicitly stated in this form for the first time by Planck, 
but it was definitively known by Boltzmann and is carved in his tombstone in Vienna.
The index~$G$ stands for Gibbs, who put Boltzmann's ideas forward 
and overspread the (classical) statistical mechanics concepts through his seminal book~\cite{Gibbs1902}.
Equation~\eqref{eq:Tombstone} is a particular instance of a more general one, namely
\begin{equation}
S_{BG} = - k\sum_{i=1}^{W} p_i \ln p_i \;\;\;\; \Bigl(\sum_{i=1}^W p_i=1 \Bigr) \,.
\label{SBG}
\end{equation}
When every microstate is equally probable, \emph{i.e.}, when $p_i=1/W\,\, \forall\, i$, we recover Eq. \eqref{eq:Tombstone}.
Evidently quantum mechanics was unknown to Boltzmann and it was just birthing when Gibbs' book was published.
It was left to von Neumann to extend Eq.~\eqref{SBG} 
in order to encompass quantum systems. 
He showed that the entropy for a quantum system should be expressed 
by using the density matrix operator~$\widehat{\rho}$, namely
\begin{equation}
S_{BG} = -k \mathrm{Tr}\comu{\widehat{\rho}\,\ln\widehat{\rho}\,}  \;\;\;(\mathrm{Tr}\widehat{\rho}=1)    \,,
\label{eq:Entropia_von_Neumann}
\end{equation}
sometimes referred to as the Boltzmann-Gibbs-von Neumann entropy (or just von Neumann entropy).
Notice indeed that the above equation recovers Eq. (\ref{SBG}) when $\widehat{\rho}$\, is diagonal.

The optimization of the entropy with appropriate constraints provides the thermal equilibrium distribution, namely the celebrated BG exponential distribution, whose consequences are consistent with classical thermodynamics.
In what follows we shall, however, see that entropic functionals different from the BG one must be used in order to satisfy thermodynamics for complex systems which strongly violate the probabilistic independence (or quasi-independence) hypothesis on which the BG entropy is generically based. This is typically the case whenever there is breakdown of ergodicity. Several dozens of non-BG entropic functionals have been studied along quite a few decades. We focus here on the following one (introduced in \cite{Tsallis1988} with the aim to generalize the BG statistical mechanics):
\begin{equation}
S_{q} = k\frac{1-\sum_{i=1}^{W} p_i^q}{q-1} =k\sum_{i=1}^W p_i \ln_q \frac{1}{p_i}  =-k\sum_{i=1}^W p_i^q \ln_q p_i =  -k\sum_{i=1}^W p_i \ln_{2-q} p_i \,,
\label{eq:SBG}
\end{equation}
where $q \in {\it R}$, and $\ln_q z \equiv \frac{z^{1-q}-1}{1-q}$ ($\ln_1 z=\ln z$). We straightforwardly verify that $\lim_{q\to 1}S_q=S_{BG}$. The inverse of the $q$-logarithmic function $\ln_q z$ is the $q$-exponential function $e_q^z \equiv [1+(1-q)z]^\frac{1}{1-q}$, if $1+(1-q)z >0$, and zero otherwise ($e_1^z=e^z$).

An entropic functional $S$ is said {\it additive} if it satisfies\cite{Penrose1970}, for any two {\it probabilistically independent} systems $A$ and $B$, that $S(A+B)=S(A) + S(B)$; otherwise it is said {\it nonadditive}. We easily verify that
\begin{equation}
\frac{S_q(A+B)}{k}=\frac{S_q(A)}{k} +\frac{S_q(B)}{k} +(1-q) \frac{S_q(A)}{k}\frac{S_q(B)}{k} \,.
\end{equation}
Therefore $S_{BG}$ is additive, and $S_q$ (with $q \ne 1$) is nonadditive. The generalization of the BG thermostatistical theory is currently referred to as nonextensive statistical mechanics \cite{Tsallis1988,Tsallis1995,GellMannTsallis2004,Tsallis2009,Tsallis2009b,Tsallis2009c,Tsallis2012,Tsallis2014} (see \cite{bibliography} for a regularly updated Bibliography. The entropy $S_q$ satisfies several interesting properties; among them, the uniqueness theorems proved by Santos\cite{Santos1997} and by Abe\cite{Abe2000}, as well as the connection \cite{TsallisHaubold2015} with the Einstein likelihood factorization principle deserve a special mention.

The natural, artificial and social complex systems to which $S_q$ and its associated statistical mechanics have been applied are very diverse. They include long-range interacting many-body Hamiltonian systems (see \cite{JundKimTsallis1995,Grigera1996,CannasTamarit1996,SampaioAlbuquerqueFortunato1997,AnteneodoTsallis1998,SilvaRegoLucenaTsallis1999,TamaritAnteneodo2000,CondatRangelLamberti2002,FulcoSilvaNobreLucena2003,AndradePinho2005,AntoniRuffo1995,LatoraRapisardaTsallis2001,CampaGiansantiMoroni2002,PluchinoLatoraRapisarda2004,MoyanoAnteneodo2006,PluchinoRapisardaTsallis2007,PluchinoRapisardaTsallis2008,ChavanisCampa2010,CampaChavanis2013,EttoumiFirpo2013,CirtoAssisTsallis2014,AntonopoulosChristodoulidi2011,ChristodoulidiTsallisBountis2014} for an overview) 
of various types and symmetries (let us incidentally mention that long-range versions of the interesting types focused on in \cite{CarideTsallisZanette1983,TsallisSilvaMendes1997} have not yet been handled), as well as non-Hamiltonian ones \cite{MiritelloPluchinoRapisarda2009}, low-dimensional dynamical systems \cite{TsallisPlastinoZheng1997,LyraTsallis1998,Lyra1998,TirnakliTsallisLyra1999,BaldovinRobledo2002,BorgesTsallisAnanosOliveira2002,AnanosTsallis2004,BaldovinRobledo2004a,BaldovinRobledo2004b,MayoralRobledo2004a,MayoralRobledo2004b,CasatiTsallisBaldovin2005,RuizTsallis2009,TirnakliBeckTsallis2007,TirnakliTsallisBeck2009,Grassberger2009,AnanosBaldovinTsallis2005,LuqueLacasaRobledo2012}, cold atoms \cite{DouglasBergaminiRenzoni2006,BagciTirnakli2009,LutzRenzoni2013},  plasmas \cite{LiuGoree2008,GhoshChatterjeeRoychoudhury2012,GhoshGhoshChatterjeeBachaTribeche2013,EmamuddinMamun2014,BouzitGougamTribeche2014,GuoMeiZhang2015,ElTantawyWazwazSchlickeiser2015,BouzitGougamTribeche2015,BouzitTribecheBains2015}, trapped atoms \cite{DeVoe2009}, spin-glasses\cite{PickupCywinskiPappasFaragoFouquet2009}, granular matter \cite{CombeRichefeuStasiakAtman2015}, high-energy particle collisions \cite{CMSetal,CleymansDeppman},
black holes and cosmology \cite{OliveiraSautuSoaresTonini1999,OliveiraSoaresTonini2004,OliveiraSoares2005,KomatsuKimura2013}, chemistry \cite{AndradeMundimMalbouisson2008,SanchezOldenhofFreitezMundim Ruette2010,SilvaAquilantiOliveiraMundim2013}, economics \cite{Borland2002,LudescherTsallisBunde2011,LudescherBunde2014}, earthquakes \cite{AntonopoulosMichasVallianatosBountis2014}, biology \cite{UpadhyayaRieuGlazierSawada2001,BogachevKayumovBunde2014}, solar wind \cite{BurlagaVinas2005,BurlagaNess2013}, anomalous diffusion and central limit theorems\cite{DrazerWioTsallis2000,MoyanoTsallisGellMann2006,Chavanis2008,UmarovTsallisSteinberg2008,UmarovTsallisGellMannSteinberg2010}, quantum entangled and nonentangled systems \cite{CarusoTsallis2008,SaguiaSarandy2010,NobreRegoMonteiroTsallis2011,NayakSudhaRajagopalUshaDevi2016}, quantum chaos \cite{WeinsteinLloydTsallis2002}, astronomical systems \cite{BetzlerBorges2012,BetzlerBorges2015}, signal and image processing \cite{GameroPlastinoTorres1997,CapurroDiambraLorenzoMacadarMartinMostaccioPlastinoRofmanTorresVelluti1998,MohanalinBeenamolKalraKumar2010,DinizMurtaBrumAraujoSantos2010,ShiLiMiaoHu2012}, self-organized criticality \cite{TamaritCannasTsallis1998}, mathematical structures \cite{NivanenLeMehauteWang2003,Borges2004,Tempesta2011,RuizTsallis2012,Touchette2013,RuizTsallis2013,Kalogeropoulos2015}, scale-free networks \cite{SoaresTsallisMarizSilva2005,ThurnerTsallis2005,MenesesCunhaSoaresSilva2006}, among others.

\section{$q$-triplets}

The optimization of $S_{BG}$ under appropriate constraints yields {\it exponential} forms for the probabilities, as well as for other relevant thermostatistical quantities (relaxation behaviors and sensitivity to the initial conditions are typical dynamical ones). It happens, however, that virtually all those natural, artificial and social systems usually considered as complex violate this behavior. Indeed they asymptotically present slower behaviors such as  (very frequently) power-laws, and (occasionally) stretched exponentials, to only mention the most typical ones\footnote{By the way let us mention a very frequent error in the literature, namely a confusion between asymptotic and strict power-law behaviors.  The original Pareto law refers to a distribution which only {\it asymptotically} behaves like a power law, say $1/x^\beta$. In fact, no nonzero distribution of a real positive  unbounded random continuous variable can exist with a single power law since it is non normalizable; indeed, $\int_0^\infty dx\, x^{-\beta}$ diverges for {\it any} real value of $\beta$. The L\'evy, $q$-exponential, $q$-Gaussian, $(q,\alpha)$-stable distributions \cite{UmarovTsallisSteinberg2008,UmarovTsallisGellMannSteinberg2010,MoyanoTsallisGellMann2006} are all different. They constitute but a few among the infinitely many distributions which asymptotically behave as a power law. They exhibit nevertheless important differences for finite values of the random variable. It is therefore a severe misuse to plainly refer, as regretfully done very frequently in the literature, to  ``L\'evy distribution" every time that in a log-log plot a straight line is observed along some decades. As said, L\'evy distributions are only one case among infinitely many which asymptotically behave as power-laws (for other similar misnames, see \cite{TsallisArenas2014}).}.

Let introduce now the $q$-triplet along the lines of \cite{Tsallis2009}. We consider the following ordinary differential equation
\begin{equation}
\frac{dy}{dx}=a\,y\;\;\;\;(y(0)=1) \,,
\label{5.13}
\end{equation}
whose solution is given by
\begin{equation}
y=e^{\,a\,x} \,.
\label{5.14}
\end{equation}
We may think of it in at least three different physical manners,  related respectively to the sensitivity $\xi$ to the initial conditions, to the relaxation in phase space (of say the BG entropy towards its value at thermal equilibrium), and, if the system is Hamiltonian, to the distribution of energies (or analogous quantities such as the distribution of velocities) at thermal equilibrium. In the first interpretation we refer to the exponential divergence with time of two trajectories in phase space with slightly different initial conditions. In the second interpretation, we focus some relaxing relevant quantity
\begin{equation}
\Omega(t) \equiv \frac{O(t)-O(\infty)}{O(0)-O(\infty)} \,,
\end{equation}
where $O$ is some dynamical observable essentially related to the evolution of the system in phase space (e.g., the time evolution of entropy while the system approaches equilibrium). We typically expect
\begin{equation}
\Omega(t) = e^{-t/\tau_1} \,,
\end{equation}
where $\tau_1$ is the relaxation time (depending on the physical property that we are focusing on, it might be $1/\tau_1 \simeq \lambda_1$ or not, where $\lambda_1$ is the maximal Lyapunov exponent). Finally, in the third interpretation, we have
\begin{equation}
Z_1p_i=e^{-\beta E_i} \,,
\end{equation}
where $Z_1\equiv \sum_{j=1}^W e^{-\beta E_j}$ is the BG partition function. The various interpretations are summarized in Table \ref{table5.1}.

\begin{table}[htbp]
\begin{center}
\begin{tabular}{c||c|c|c||}
                                                            &  $x$         & $a$                            &$y(x)$                                           \\
[1mm] \hline\hline 
& & & \\
Equilibrium distribution                        &  $E_i$     & $-\beta$                     & $Z_1p(E_i)=e^{-\beta E_i}$        \\
[3mm] \hline
& & & \\
Sensitivity to the initial conditions       &  $t$         & $\lambda_1$              & \;\;\;\;\;$\xi(t)=e^{ \, \lambda_1\,t}$             \\
[3mm] \hline 
& & & \\
Typical relaxation of observable $O$   &$t$           &$-1/\tau_1$                &\;\;\;\;\;$\Omega(t)=e^{-t/\tau_1}$             \\
[3mm] \hline \hline
\end{tabular}
\end{center}
\caption{Three possible physical interpretations of Eq. (\ref{5.14}) within $BG$ statistical mechanics.}
\label{table5.1}
\end{table}

Let us now generalize these statements. The solution of the diferential equation
\begin{equation}
\frac{dy}{dx}=a\,y^q\;\;\;\;(y(0)=1)
\label{5.18}
\end{equation}
is given by
\begin{equation}
y= [1+(1-q)a\,x]^{\frac{1}{1-q}} \equiv e_q^{\,a\,x} \,.
\label{5.19}
\end{equation}
These expressions respectively generalize expressions (\ref{5.13}) and (\ref{5.14}). As before,
we may think of them in three different physical manners,  related respectively to the sensitivity to the initial conditions, to the relaxation in phase space, and, if the system is Hamiltonian, to the distribution of energies at a stationary state. In the first interpretation we reproduce $\xi=e_{q_{sen}}^{\lambda_{q_{sen}} \,t}$. In the second interpretation, we typically expect
\begin{equation}
\Omega(t) = e_{q_{rel}}^{-t/\tau_{q_{rel}}} \,,
\end{equation}
where $\tau_{q_{rel}}$ is the relaxation time. Finally, in the third interpretation, we have 
\begin{equation}
Z_{q_{stat}}p_i=e^{-\beta_{q_{stat}} E_i} \,,
\end{equation}
where $Z_{q_{stat}}\equiv \sum_{j=1}^W e_{q_{stat}}^{-\beta_{q_{stat}} E_j}$ is the $q$-generalized partition function. The various interpretations are summarized in Table \ref{table5.2}. The set $(q_{sen},q_{rel},q_{stat})$ constitutes what we shall refer to as the {\it $q$-triplet} (occasionally referred also to as the {\it $q$-triangle}). In the $BG$ particular case, we recover $q_{sen}=q_{rel}=q_{stat}=1$. The existence of these three $q$-exponentials characterized by the $q$-triplet was predicted in 2004 \cite{Tsallis2004}, and confirmed in 2005 \cite{BurlagaVinas2005} in the solar wind (by processing the data sent to Earth by the spacecraft Voyager 1); more along these lines can be found in \cite{LeubnerVoros2005a,LeubnerVoros2005b}.

\begin{table}[htbp]
\begin{center}
\begin{tabular}{c||c|c|c||}
                                                            &  $x$         & $a$                                       &$y(x)$                                                                             \\
[1mm] \hline\hline
& & & \\
Stationary state distribution                &  $E_i$     & $-\beta$                                 & $Z_{q_{stat}} \, p(E_i)=e_{q_{stat}}^{-\beta E_i}$        \\
[3mm] \hline
& & & \\
Sensitivity to the initial conditions       &  $t$         & $\lambda_{q_{sen}}$              & \;\;\;\;\;\;\;\;\;\;\;\;\;\;\;$\xi(t)=e_{q_{sen}}^{ \, \lambda_{q_{sen}}\,t}$             \\
[3mm] \hline 
& & & \\
Typical relaxation of observable $O$   &$t$           &$-1/\tau_{q_{rel}}$                  & \;\;\;\;\;\;\;\;\;\;\;\;\;\;\;$\Omega(t)=e_{q_{rel}}^{-t/\tau_{q_{rel}}}$                   \\
[3mm] \hline \hline
\end{tabular}
\end{center}
\caption{Three possible physical interpretations of Eq. (\ref{5.19}) within nonextensive statistical mechanics.}
\label{table5.2}
\end{table}

A plethora of $q$-triplets have been found in solar plasma \cite{BurlagaVinas2005,BurlagaNess2013,PavlosKarakatsanisXenakis2012,KarakatsanisPavlosXenakis2013,PavlosIliopoulosZastenkerZelenyiKarakatsanisRiazantsevaXenakisPavlos2015}, the ozone layer \cite{FerriSavioPlastino2010}, logistic map (see \cite{TsallisPlastinoZheng1997,LyraTsallis1998,Lyra1998,BaldovinRobledo2004a,BaldovinRobledo2004b,MayoralRobledo2004a,MayoralRobledo2004b,TirnakliBeckTsallis2007,TirnakliTsallisBeck2009,Grassberger2009,AnanosBaldovinTsallis2005,LuqueLacasaRobledo2012}), El Ni\~no/Southern Oscillation \cite{FerriFigliolaRosso2012}, geological faults \cite {FreitasFrancaScherrerVilarSilva2013}, finance\cite{PavlosKarakatsanisXenakisPavlosIliopoulosSarafopoulos2014,IliopoulosPavlosMagafasKarakatsanisXenakisPavlos2015}, DNA sequence \cite{PavlosKarakatsanisIliopoulosPavlosXenakisClarkDukeMonos2015}, and elsewhere \cite{Tsallis2006,SuyariWada2007}.

\section{Generalizing the additive and multiplicative self-dual relations}
\label{Sec:Generalizing}

\begin{figure}[h!]
\begin{center}
\includegraphics[width=0.8\columnwidth,angle=0]{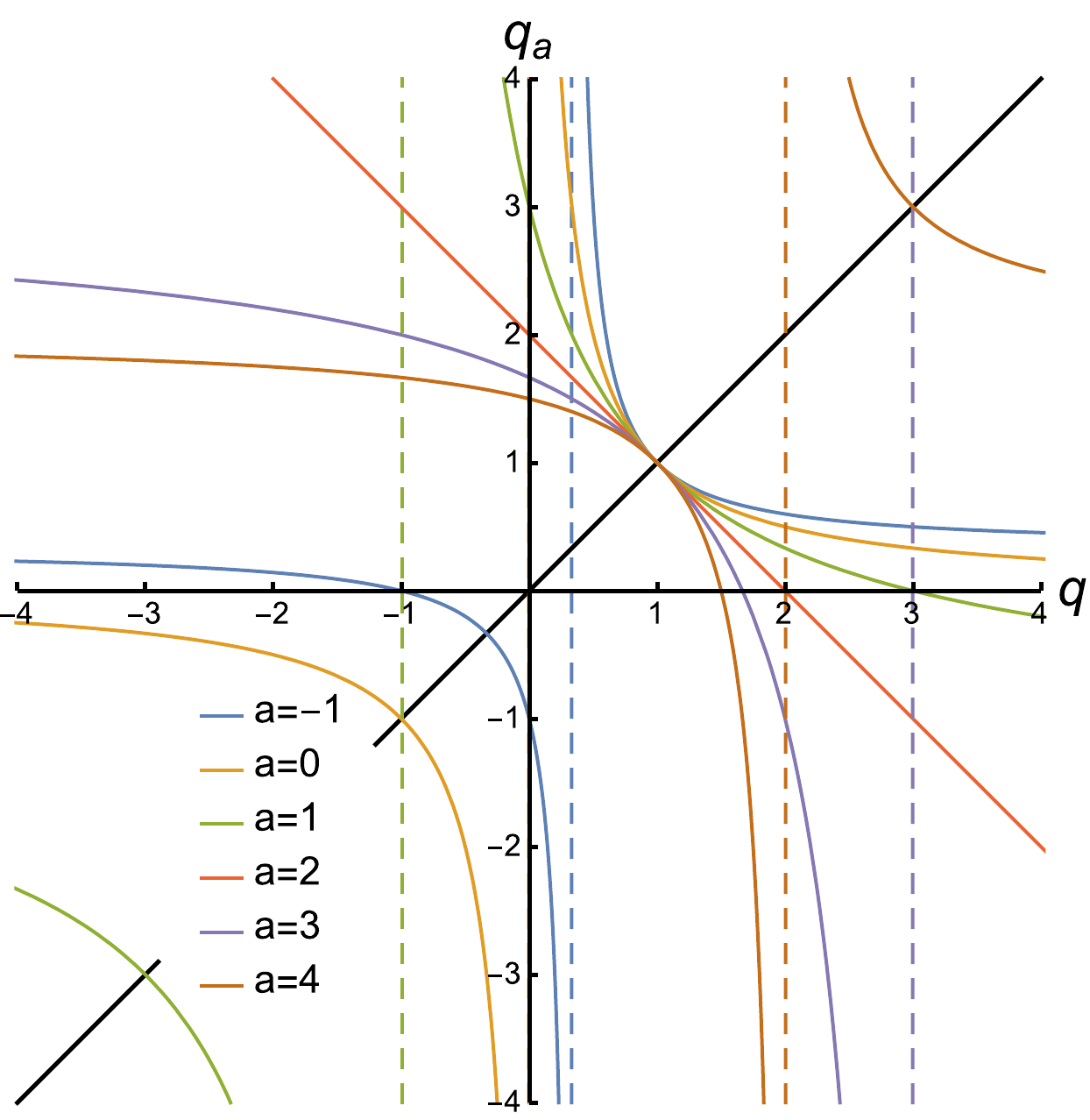}
\end{center}
\caption{The self-dual transformation $q_a(q)$ as given by Eq. (\ref{qdualitynew}), for typical values of $a$; $q_2(q)=2-q$ recovers the additive duality; $q_0(q)=1/q$ recovers the multiplicative duality. For $a>2$, when $q$ varies within $(-\infty,1]$, $q_a$ varies biunivocally within $[\frac{a}{a-2},1]$, when $q$ varies within $[1,\frac{a}{a-2}]$, $q_a$ varies biunivocally within $[1,-\infty)$, and when $q$ varies within $[\frac{a}{a-2},\infty)$, $q_a$ varies biunivocally within $[\infty,\frac{a}{a-2}]$. For $a<2$, when $q$ varies within $(-\infty,\frac{a}{-(2-a)}]$, $q_a$ varies biunivocally within $[\frac{a}{-(2-a)},-\infty)$, when $q$ varies within $[\frac{a}{-(2-a)},1]$, $q_a$ varies biunivocally within $(\infty,1]$, and when $q$ varies within $[1,\infty]$, $q_a$ varies biunivocally within  $[1,\frac{a}{-(2-a)}]$.}
\label{selfduality}
\end{figure}

Let us consider the following transformation:
\begin{equation}
q_a=\frac{(a+2) -aq}{a-(a-2)q} \,,
\label{qdualitynew}
\end{equation}
or, equivalently,
\begin{equation}
\frac{1}{1-q_a}=\frac{1}{q-1}  + 1-\frac{a}{2}\,.
\label{qduality2}
\end{equation}
We straightforwardly verify that $q_2=2-q$ ({\it additive duality}) and $q_0=1/q$ ({\it multiplicative duality}) \cite{TsallisGellMannSato2005,Tsallis2009,Tsallis2009b,Tsallis2012}. Also, we generically verify {\it selfduality}, i.e., $q_a(q_a(q))=q \,, \forall (a,q)$, as well as the BG fixed point, i.e., $q_a(1)=1 \,, \forall a$: See Fig. \ref{selfduality}. The duality \eqref{qdualitynew} is in fact the most general ratio of linear functions of $q$  which satisfies these two important properties (selfduality and BG fixed point). It transforms biunivocally the interval $[1,-\infty)$ into the interval $[1,\frac{a}{a-2}]$.
Moreover, for $a=3$ and $a=5$ we recover respectively $q_3=\frac{5-3q}{3-q}$ \cite{NelsonUmarov2010} and $q_5=\frac{7-5q}{5-3q}$ \cite{HanelThurnerTsallis2009}. 

Let us combine now two\footnote{It is also possible to combine, along similar lines, three or more such transformations.} transformations of the type  \eqref{qdualitynew} (or, equivalently, \eqref{qduality2}):
\begin{equation}
\mu \; \to \; q_a(q)=\frac{(a+2) -aq}{a-(a-2)q} \; \to \; \frac{1}{1-q_a(q)}=\frac{1}{q-1}  + 1-\frac{a}{2} \,,
\label{transform1}
\end{equation}
and
\begin{equation}
\nu \; \to \; q_b(q)=\frac{(b+2) -bq}{b-(b-2)q} \; \to \; \frac{1}{1-q_b(q)}=\frac{1}{q-1}  + 1-\frac{b}{2} \,,
\label{transform2}
\end{equation}
with $b \ne a$. It follows that
\begin{equation}
\mu \nu  \; \to \; q_a(q_b(q))=\frac{(b-a) -(b-a-2)q}{(b-a+2)-(b-a)q} \; \to \; \frac{1}{1-q_a(q_b(q))}=\frac{1}{1-q}  + \frac{b-a}{2} \,,
\label{transform3}
\end{equation}
and
\begin{equation}
\nu \mu  \; \to \; q_b(q_a(q))=  \frac{(a-b) -(a-b-2)q}{(a-b+2)-(a-b)q} \; \to \; \frac{1}{1-q_b(q_a(q))}=\frac{1}{1-q}  + \frac{a-b}{2} \,,
\label{transform4}
\end{equation}
with $\mu^2=\nu^2=1$, $\nu \mu=(\mu \nu)^{-1}$, and $q_a(q_a(q))=q \,, \forall (a,q)$. 

For integer values of $m$ and $n$, we can straightforwardly establish
\begin{eqnarray}
(\mu \nu)^m  \; &\to& \; q_{a,b}^{(m)}(q) \equiv q_a(q_b(q_a(q_b(...))))=\frac{m(b-a) -[m(b-a)-2]q}{[m(b-a)+2]-m(b-a)q} \\ \; &\to& \; \frac{1}{1-q_{a,b}^{(m)}(q)} =\frac{1}{1-q_a(q_b(q_a(q_b(...))))}=\frac{1}{1-q}  + m\frac{b-a}{2} \,,
\label{algebra1}
\end{eqnarray}
and
\begin{eqnarray}
(\nu \mu)^n  \; &\to& \; q_{b,a}^{(n)}(q) \equiv q_b(q_a(q_b(q_a(...))))=  \frac{n(a-b) -[n(a-b)-2]q}{[n(a-b)+2]-n(a-b)q} \\ \; &\to& \; \frac{1}{1-q_{b,a}^{(n)}(q)} = \frac{1}{1-q_b(q_a(q_b(q_a(...))))}=\frac{1}{1-q}  +n \frac{a-b}{2} \,.
\label{algebra2}
\end{eqnarray}
As we see, $q_{a,b}^{(1)}=q_a(q_b(q))$ and $q_{b,a}^{(1)}=q_b(q_a(q))$.

For $a \ne b$ and any integer values for $(m,n)$, the above general relations can be conveniently rewritten as follows:
\begin{eqnarray}
\frac{2}{b-a} \frac{1}{1-q_{a,b}^{(m)}(q)} =\frac{2}{b-a}\frac{1}{1-q}  + m \;\;\;(m=0,\pm 1,\pm2, ...) \,,
\end{eqnarray}
and
\begin{eqnarray}
\frac{2}{a-b} \frac{1}{1-q_{b,a}^{(n)}(q)} =\frac{2}{a-b}\frac{1}{1-q}  + n \;\;\;(n=0,\pm 1,\pm2, ...) \,.
\end{eqnarray}
For $m=n=1$ and $(a,b)=(2,0)$ we recover the simple transformations $q_{2,0}^{(1)}=2-\frac{1}{q}$ (see Eq. (7) in \cite{MoyanoTsallisGellMann2006}, and footnote in page 15378 of \cite{TsallisGellMannSato2005}) and $q_{0,2}^{(1)}=\frac{1}{2-q}$.

We can also check that, with $m=0,\pm 1,\pm 2, ...$, $(\mu \nu)^m \mu$ and $\nu(\mu \nu)^m$ correspond respectively to 
\begin{eqnarray}
\frac{2}{b-a} \frac{1}{1-q_{a,b}^{(m, \mu)}(q)} -\frac{2-a}{2(b-a)}=- \Bigl[\frac{2}{b-a} \frac{1}{1-q} -\frac{2-a}{2(b-a)} \Bigr] - m \,,
\label{algebra3}
\end{eqnarray}
and
\begin{eqnarray}
\frac{2}{b-a} \frac{1}{1-q_{a,b}^{(\nu, m)}(q)} -\frac{2-b}{2(b-a)}=  -\Bigl[\frac{2}{b-a} \frac{1}{1-q} -\frac{2-b}{2(b-a)}\Bigr] + m \,.
\label{algebra4}
\end{eqnarray}

Analogously we can check that, with $n=0,\pm 1,\pm 2, ...$, $(\nu \mu)^n\nu$ and  $\mu(\nu \mu)^n$ correspond respectively to
\begin{eqnarray}
\frac{2}{a-b} \frac{1}{1-q_{b,a}^{(n, \nu)}(q)} -\frac{2-b}{2(a-b)}=- \Bigl[\frac{2}{a-b} \frac{1}{1-q} -\frac{2-b}{2(a-b)} \Bigr] - n \,,
\label{algebra5}
\end{eqnarray}
and
\begin{eqnarray}
\frac{2}{a-b} \frac{1}{1-q_{b,a}^{(\mu, n)}(q)} -\frac{2-a}{2(a-b)}=  -\Bigl[\frac{2}{b-a} \frac{1}{1-q} -\frac{2-a}{2(a-b)}\Bigr] + n \,.
\label{algebra6}
\end{eqnarray}

As we see, the algebras that are involved exhibit some degree of complexity. Let us therefore summarize the frame within which we are working. If we have an unique parameter (noted $a$) to play with, we can only transform $q$ through Eq. (\ref{qdualitynew}). If we have two parameters (noted $a$ and $b$) to play with, we can transform $q$ in several ways, namely through Eqs. (\ref{algebra1}), (\ref{algebra2}), (\ref{algebra3}), (\ref{algebra4}), (\ref{algebra5}) and (\ref{algebra6}), with $m=0,\pm 1,\pm 2, ...$ and $n=0,\pm 1,\pm 2, ...$; the cases $m=0$ and $n=0$ recover respectively Eqs. (\ref{transform1}) and (\ref{transform2}). The particular choice $(a,b)=(2,0)$ yields the algebra introduced in \cite{TsallisGellMannSato2005,Tsallis2009,Tsallis2009b,Tsallis2012}. Also, the particular choice $(a,b)=(-1,0)$ within the transformation (\ref{transform3}) recovers the transformation $q \to \frac{1+q}{3-q}$, which plays a crucial role in the $q$-generalized Central Limit Theorem \cite{UmarovTsallisSteinberg2008}. More generally, the relation $b-a=1$ recovers the $\gamma=1/2$ case of Eq. (32) of \cite{Tsallis2016} (see also \cite{RuizTsallis2012,Touchette2013,RuizTsallis2013}).

\section{Some final remarks}

The data observed in \cite{BurlagaVinas2005} for the solar wind are consistent with the $q$-triplet \cite{TsallisGellMannSato2005} $(q_{sen}, q_{stat},q_{rel})=(-0.5,7/4,4)$.

If we identify, in Eq. (\ref{transform3}), $(q,q_{a,b}^{(1)}) \equiv (q_{sen},q_{rel})$ we can verify that, for $a-b=2$, the data are consistently recovered. Moreover, if we use once again Eq. (\ref{transform3}) and $a-b=2$, but identifying now $(q,q_{a,b}^{(1)}) \equiv (q_{rel},q_{stat})$, once again the data are consistently recovered. The particular case $(a,b)=(2,0)$ was first proposed in \cite{TsallisGellMannSato2005}. In other words, it is possible to consider this $q$-triplet as having only one independent value, say $q_{sen}$; from this value we can calculate $q_{rel}$ by using  Eq. (\ref{transform3}); and from $q_{rel}$ we can calculate $q_{stat}$ by using once again Eq. (\ref{transform3}). This discussion can be summarized as follows:
\begin{equation}
\frac{1}{1-q_{sen}}-\frac{1}{1-q_{rel}}  = \frac{1}{1-q_{rel}}-\frac{1}{1-q_{stat}} =   \frac{a-b}{2}=1 \,.
\end{equation}
It is occasionally convenient to use the $\epsilon$-triplet defined as $(\epsilon_{sen},\epsilon_{stat},\epsilon_{rel})= (1-q_{sen},1-q_{stat},1-q_{rel})$. Let us mention that an amazing set of relations was found among these by \cite{Baella2008}, namely 
\begin{eqnarray}
\epsilon_{stat} = \frac{\epsilon_{sen} + \epsilon_{rel}}{2} \,,    \\
\epsilon_{sen}=\sqrt{\epsilon_{stat} \, \epsilon_{rel}}  \,, \\
\epsilon_{rel}^{-1} =\frac{\epsilon_{sen}^{-1} + \epsilon_{stat}^{-1}}{2}   \,.
\end{eqnarray}
The emergence of the three Pythagorean means in this specific $q$-triplet remains still today enigmatic.

Let us now focus on a different system, namely the well known logistic map at its edge of chaos (also referred to as the Feigenbaum point). The numerical data for this map yield the $q$-triplet  $(q_{sen}, q_{stat},q_{rel})=(0.244487701...,1.65 \pm 0.05,2.249784109...)$ \cite{LyraTsallis1998,MouraTirnakliLyra2000,Grassberger2005,Robledo2006,TirnakliTsallisBeck2009}.

An heuristic relation has been found \cite{Baella2010} between these three values, namely 
(using $\epsilon \equiv 1-q$)
\begin{equation}
\epsilon_{sen} + \epsilon_{rel} = \epsilon_{sen} \, \epsilon_{stat} \,.
\label{baella2}
\end{equation}
Indeed, this relation straightforwardly implies
\begin{equation}
q_{stat}=\frac{q_{rel}-1}{1-q_{sen}} \,.
\end{equation}
Through this relation we obtain $q_{stat}=1.65424...$ which is perfectly compatible with $1.65 \pm 0.05$. 
In the generalized algebra that we have developed here above we have three free parameters $(q,a,b)$ in addition to the integer numbers $(m,n)$. It is therefore trivial to make analytical identifications with $(q_{sen}, q_{stat},q_{rel})$ such that Eq. (\ref{baella2}) is satisfied. 

The real challenge, however, is to find a general theoretical frame within which such identifications (and, through the freedom associated with $(m,n)$, infinitely many more, related to physical quantities) become established on a clear basis, and not only through conjectural possibilities. 
Such a frame of systematic identifications remains up to now elusive and certainly constitutes a most interesting open question. Along this line, a connection that might reveal promising is that, if we assume that $q$ is a complex number (see, for instance, \cite{WilkWlodarczyk2015,AzmiCleymans2015}), then Eq. (\ref{qdualitynew}) corresponds to nonsingular [with $(a+2)(a-2)-a^2=-4 \ne 0\,,\forall a$] Moebius transformations, which form the Moebius group, defining an automorphism of the Riemann sphere.

\subsection*{Acknowledgments}
I have benefited from fruitful remarks by D. Bagchi, E.M.F. Curado, A.R. Plastino, P. Rapcan, G. Sicuro, P. Tempesta and U. Tirnakli.
I also acknowledge partial financial support from CNPq and Faperj (Brazilian agencies), as well as from the John Templeton Foundation (USA).
It is both a great pleasure and a honor to dedicate this manuscript to my friend \textbf{Alberto Robledo},
wishing him a very happy beginning for his second 70 years!

\end{document}